\RequirePackage{fix-cm}
\documentclass[smallextended]{svjour3}       
\smartqed  
\usepackage{graphicx}
\usepackage{ucs}
\usepackage[utf8x]{inputenc}
\usepackage{
	amsmath,
	amssymb,
	amsfonts,
	fancyhdr,
	graphicx,
	subfigure,
	color
	}
\usepackage[locale=DE]{siunitx}	
\usepackage{booktabs}
\usepackage{longtable}
\usepackage{tabularx}
\usepackage{hyperref}
\usepackage{braket}
\usepackage{varwidth} 
 \newcommand{\bvec}[1]{\boldsymbol{#1}}
\newcommand{\bk}[1]{\left(#1\right)}					

\begin{document}

\title{Transport and quantum coherence in graphene rings}
\subtitle{Aharonov-Bohm oscillations, Klein tunneling and particle localization} 


\titlerunning{ Transport and quantum coherence in graphene rings}        

\author{Alexander Filusch \and Christian Wurl \and
        Andreas Pieper \and
        Holger Fehske 
}


\institute{A. Filusch \at Institute of Physics, University Greifswald, 17487 Greifswald, Germany\\
              Tel.: +49 3834 420 4724\\
              \email{af142713@uni-greifswald.de}           
\and C. Wurl \at Institute of Physics, University Greifswald, 17487 Greifswald, Germany\\
              Tel.: +49 3834 420 4724\\
              \email{wurl@physik.uni-greifswald.de}           
\and A. Pieper \at Institute of Physics, University Greifswald, 17487 Greifswald, Germany\\
              Tel.: +49 3834 420 4768\\
              \email{pieper@physik.uni-greifswald.de}           
\and H. Fehske (corresponding autor) \at Institute of Physics, University Greifswald, 17487 Greifswald, Germany\\
              Tel.: +49 3834 420 4760\\
              \email{fehske@physik.uni-greifswald.de}           
}

\date{Received: date / Accepted: date}

\maketitle

\begin{abstract}
Simulating quantum transport through  mesoscopic, ring-shaped graphene structures, we address various quantum coherence and interference phenomena. First,  a perpendicular magnetic field, penetrating the graphene ring, 
gives rise to Aharonov-Bohm oscillations in the conductance as a function of the magnetic flux, on top of the universal conductance fluctuations. At very high fluxes the interference gets suppressed and quantum Hall edge channels develop. Second, applying an electrostatic potential to one of the ring arms, $nn'n$- or $npn$-junctions can be  realized with  particle transmission due to normal tunneling or Klein tunneling. In the latter case 
the Aharonov-Bohm oscillations weaken for smooth barriers. Third, if potential disorder comes in to play, both Aharonov-Bohm and Klein tunneling effects rate down, up to the point where particle localization sets in.  
\keywords{mesocopic transport \and quantum interferece  \and graphene-based nanostructures \and Aharonov-Bohm effect \and disorder effects}
 \PACS{73.23.-b \and 72.80.Vp \and 73.43.Jn \and 73.20.Fz}
\end{abstract}

\section{Introduction}
\label{intro}
Quantum coherence and interference effects are fundamental for the description of transport in mesoscopic devices~\cite{Dat95}. Graphene, a strictly two-dimensional material with honeycomb lattice structure that causes the nontrivial topology of the electronic wave function and an almost linear low-energy spectrum of the chiral quasiparticles (charge carriers) near the so-called Dirac nodal points, has opened new perspectives for mesoscopic physics~\cite{CNG09}.  Thereby, from an application-technological point of view, the tunability of the transport properties of graphene-based nanostructures by external fields--allowing a controlled modification of selected areas of the sample by gating--is of particular importance~\cite{Go11}. 

The occurrence of so-called Aharonov-Bohm (AB) oscillations~\cite{AB59} in the conductance of ring-shaped graphene devices, measured as a function of a magnetic field applied perpendicular to the graphene plane in a two-terminal setup, is perhaps the most basic and direct particle interference effect observed in graphene systems so far~\cite{RO08}. 
The experiments reveal clear oscillations of the magneto-conductance with a period corresponding to one flux quantum $\Phi_0=h/e$, on top of the universal conductance fluctuations that always occur in such mesoscopic devices. The experimental results were confirmed with improved resolution for smaller ring structures~\cite{HMJPSEI09}, even up to third harmonics of the AB oscillations~\cite{Naea12}. Also the influence of  in-plane gates was studied in detail~\cite{HMJPSEI10}.  Since in graphene both charge carrier types, electrons and holes, can be induced in one and the same sample with local gates, a ring geometry allows to study not only the quantum interference of electrons (holes) with electrons (holes) but also between electrons and holes. In fact, the AB effect has  been found in an electron-hole graphene ring system~\cite{SSH12}. On the theoretical side, starting with the pioneering work~\cite{Reea07} on the AB effect in isolated circular and hexagonal rings, a great variety of topics has been addressed,  based on both continuum~\cite{JMPT09,SZ16} and tight-binding~\cite{SB10,WW10,KKM17} model descriptions (For an overview, also of a wider class of AB-like effects in graphene structures, see~\cite{SRT12,Kat12}). In particular signatures of valley polarization in the magneto-conductance  of a graphene AB interferometer have been identified~\cite{Ryc09}. To date, however, it has not been possible to verify most of the  graphene-specific theoretical predictions in real experiments. 

Promising in this respect seems to be a proposal~\cite{SB10,Ryc09} to tune the band structure in one of the arms of the graphene ring through the Dirac point by an electrostatic potential $V$, such that a transition from an $nn'n$-- to an $npn$--junction is induced. (Here the two adjacent leads show  $n$ conduction-band transport  if $E>0$, whereas, in the gated arm,  $n'$  conduction-band transport takes place for $E>V$ due to normal tunneling processes or $p$ valence-band transport happens for $V>E>0$ due to Klein tunneling~\cite{Kl28}.) Klein tunneling in graphene~\cite{KNG06,SHG09,NKPKKL11}, i.e., the perfect transmission of particle waves through sharp potential barriers (or $pn$--junctions) of arbitrary height and width at perpendicular incidence, is a consequence of the pseudo-relativistic dynamics of the massless chiral Dirac-Weyl quasiparticles having an additional pseudospin degree of freedom~\cite{Di28,We29}.  Accordingly, a very sharp $pn$--junction in the lower arm will not affect the AB oscillations very much, but smooth $pn$--junctions, where particles have to tunnel through a finite region of low density of states, will significantly suppress the AB signal; a graphene-specific effect that should be readily observable~\cite{SB10}.    
 
Regrettably transport through graphene-based devices is strongly affected by disorder. For example, edge roughness, intrinsic impurities, bulk defects induced by the substrate, or adatoms on the open surface of graphene lead to a strong particle scattering. Disorder is known to be exceedingly efficient in suppressing the mobility of the charge carrier  in low-dimensional systems, even to the point of Anderson localization~\cite{An58}.  However graphene shows distinctive features in this respect too~\cite{MKFSAA06}. First, only short-range impurities may cause the intervalley scattering that gives rise to particle localization.  Second, due to the chirality of the charge carriers quantum interference may trigger even (weak) antilocalization~\cite{TKSG09}. Third, charge carrier density fluctuations may break up the sample into electron-hole puddles~\cite{ACFD08,SF12a}; as a  result mesoscopic transport is rather determined  by activated hopping or leakage between the puddles. Considering the above mentioned graphene AB ring, disorder will have a lasting effect on the phase relations, i.e., the quantum interference, of particles passing through the two arms of the AB ``interferometer''.  

For these reasons it seems appropriate to extend previous work on contacted graphene rings~\cite{SB10,SRT12}, and analyze  in detail not only the interplay of Aharonov-Bohm effect and Klein tunneling, but also the influence of (bulk) disorder, which is known to be  especially important in such low-dimensional restricted geometries. We thereby pursue a microscopic approach and use exact numerical techniques  to provide unbiased data for various transport quantities. The remaining part of this paper is organized as follows. In Sect.~\ref{s_tm} we introduce the two-terminal geometry for AB rings made of graphene, specify the corresponding tight-binding Hamiltonian, and outline the theoretical (scattering-matrix based) approach. Section~\ref{results} provides our numerical results. Here, using the {\it Kwant} software package~\cite{GW14} and the Kernel polynomial method~\cite{WWAF06} we determine the transmission of (Dirac) electrons through the AB interferometer in different transport regimes, mainly in terms of the conductance and the local density of states. Compared to~\cite{SB10,SRT12} we study various  magnetic-field configurations and discuss the influence of the Lorentz force (for weak and strong fields) from a semiclassical point of view.  Finally, we conclude in  Sect.~\ref{conclus}.

\section{Theoretical modeling}
\label{s_tm}
\begin{figure}
\centering
\includegraphics[scale=0.55]{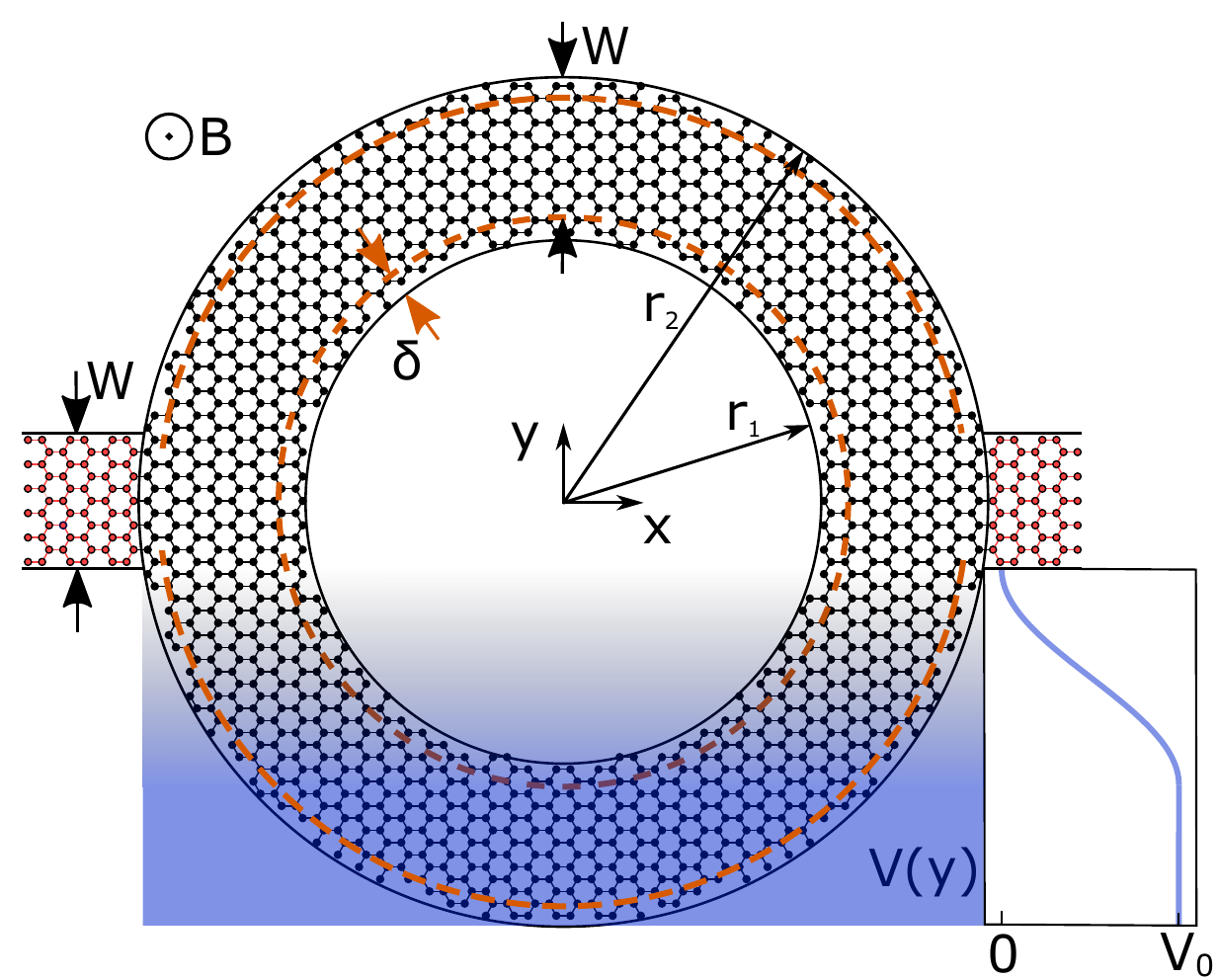}
\includegraphics[scale=0.47]{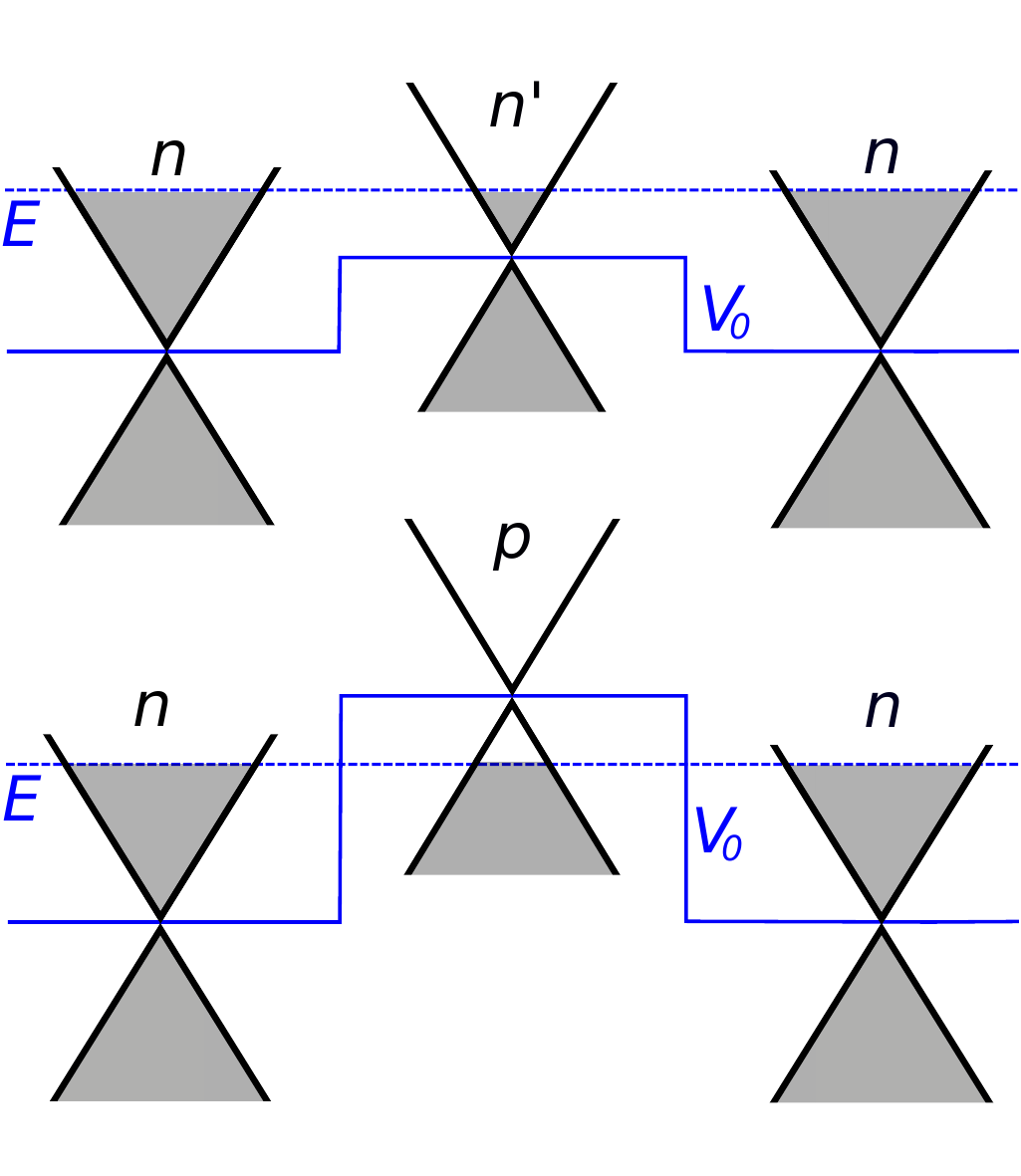}
\caption{Two-terminal setup considered in this work. Left part:  Planar graphene ring structure (black) contacted by leads (red). Besides the homogeneous magnetic field $B$ (pointing out of the plane), the lower arm of the ring is subject to  a gate potential $V$, having the pictured smooth density profile in (negative) $y$-direction while being constant along $x$-direction. The average width of the ring-shaped structure $(r_2-r_1)$ is assumed to be $(W+\delta)$ and therefore slightly wider than the width of the leads $W$. In the peripheral zones of width $\delta$ (between the black and dashed orange circles), edge disorder appears due to random on-site potentials. Throughout this work we use $r_2/a_0=300$, $W/a_0=60$, and $\delta/a_0=1.5$, where $a_0\simeq 1.42$\AA$\,$  is the distance between neighboring carbon atoms in graphene. In this case the ring contains about 80.000 lattice sites, and hence can be considered as a mesoscopic object. Right part: Schematic band structure of graphene (in continuum approximation, near the Dirac points) when an electrostatic potential $V_{0}$ acts on the lower arm of the AB interferometer.  
If the Fermi energy $E$ of the Dirac electrons is smaller (larger) than the potential barrier, $E<V_{0}$ shown on the bottom ($E>V_{0}$, top), an $npn$- ($nn'n$-) junction forms, giving rise to Klein (normal) tunneling processes.}
\label{fig1}
\end{figure}

As a  starting point for  our theoretical considerations  we assume that a ring-shaped structure together with contacts can  be cut out of a graphene sheet, see Fig.~\ref{fig1}. The inner and outer edges of the round graphene structure are irregular and were defined in a way to best approach the circular geometry. In this case, for isolated rings, one does not observe the formation of edge-state subbands with well defined number of energy levels, but in the low-flux limit of the ring spectra anti-crossing levels evolve, indicating the coupling of inner and outer edge states, just as for more regular hexagonal or rhombus-shaped graphene rings~\cite{BPS09}. The structure is exposed to a magnetic field ${\bf B}({\bf r})$ perpendicular to the graphene surface and to a local gate $V({\bf r})$ acting on the lower arm only.  

We describe the electronic properties of the system by the usual graphene tight-binding Hamiltonian~\cite{W47,CNG09} 
\begin{align}
H=- t_0 \sum_{<i,j>} {\rm e}^{\frac{{\rm i} 2\pi}{\Phi_0} \int_{\boldsymbol{r}_i}^{\boldsymbol{r}_j} \boldsymbol{A}(\boldsymbol{r}) \text{d}\boldsymbol{r}} \; c^{\dagger}_i c_j^{} + \sum_{i} [V(\boldsymbol{r}) +\Delta^{(b,e)}(\boldsymbol{r})] \,c^{\dagger}_i c_i^{}\, , \label{eq1}
\end{align}
where $c_{i}^{\dagger}$ ($c_{j}^{}$) creates (annihilates) an electron in the carbon $2p_z$ orbital at lattice site $i$ ($j$),  
and the sum in the first term runs independently over (nearest-neighbor) sites $i$ and $j$. The applied magnetic field is included by means of the Peierls  substitution~\cite{P33}, which adds a complex phase -- proportional to the magnetic flux -- to the hopping integral $t_0$. For graphene, we have $t_0\simeq 2.7$~eV. Here, the line integral over the vector potential ${\bf A}({\bf r})$ is taken along the straight path between sites $i$ and $j$. The second term takes into account the shift of the on-site energy due to gate electrode potential $V(\boldsymbol{r})$ on the lower arm, parametrized as a smooth cosine step function, 
\begin{align}
 V(y)= \frac{V_0}{2}\left[1- \cos{\Big(\frac{y+W/2}{r_2-3W/2} \cdot \pi \Big)}\right] \;\mbox{for}\;\; -r_1-\delta<y<-W/2\;,
\label{potential}
\end{align} 
while being constant $V(y)=0$ for $y\geq -W/2$ and $V(y)=V_0$ for $y\leq -r_1-\delta$. In addition, we allow for on-site 
bulk and edge disorder~\cite{SSF09}, where $\Delta^{(b)}=\Delta_i$ for $y\leq -W/2$ and $\Delta^{(e)}=\Delta_i$ for $r_1<r<r_1+\delta$ or  $r_2-\delta<r<r_2$ with  $|y|>W/2$, respectively. The random on-site potentials are drawn from the box distribution 
\begin{align}
p[\Delta_i]=\frac{1}{\gamma_{b,e}} \theta (\gamma_{b,e}/2-|\Delta_i|)\,, 
 \end{align} 
 where $\gamma_{b,e}$ measures the (bulk,edge) disorder strength.

For the actual quantum transport calculation, we use the \textit{Python}-based open-source toolbox {\it Kwant}~\cite{GW14}, where our setup contains a scattering region (ring) sandwiched between two semi-infinite graphene nanoribbons (leads).   The scattering matrix $S_{n,m}$ relates the amplitudes of the incoming waves $a_m$ in the left lead (L) to the amplitudes of outgoing (transmitted) waves $b_n$ in the right lead (R) for all open (i.e., active) channels~\cite{Dat95}. Then, in the limit of vanishing bias voltage, the transmission between the left and right leads is given within the Landauer-B\"uttiker approach~\cite{Dat95,La70,Bu86,PSWF13} as 
\begin{equation}
 T=\sum_{m \in L, n \in R} | S_{n,m} |^2\;.
\label{lb_g}
\end{equation} 
We furthermore exploit the local density of states (LDOS) at site $i$ of a given sample, 
\begin{align}
\text{LDOS}\bk{E}_i=\sum \limits_l |\langle i | l \rangle|^2 \delta\bk{E-E_l},
\label{eq:ldos}
\end{align}
as a probe of particle localization~\cite{SSF09,SB14,FHP15}.  In Eq.~(\ref{eq:ldos}), $E$ is the energy of the particle,
and the sum extends over all single-electron eigenstates $|l\rangle=c_l^\dagger |0\rangle$ of $H$ with energy $E_l$.  The LDOS can be obtained numerically by the kernel polynomial method in a very efficient way~\cite{WWAF06}. 

\section{Numerical results}
\label{results}
\subsection{Aharonov-Bohm  conductance oscillations}
\label{aboszi}
Let us start by investigating the transmission  $T$ through our graphene ring under the influence of a perpendicular magnetic field only [that means, we set $V_0=0$ and $\Delta^{(b,e)}=0$ in Eq.~(\ref{eq1})]. According to the AB effect~\cite{AB59} the magneto-conductance  $G = (2 {\rm e}^2/h) T$ (here the factor two comes from the spin degree of freedom) should oscillate as a function of the magnetic flux $\Phi$ through the ring, simply because of the phase difference  between electrons traveling along the upper respectively lower arm of the ring. As electrons can circulate around the ring several times before escaping into the right lead, $G(\Phi)$ is expected to oscillate with periods $\Delta\Phi=\Phi_0/n$ with $n \in \mathbb{N}\backslash\{0\}$. 

\begin{figure}[t!]
\centering
\includegraphics[width=1\textwidth]{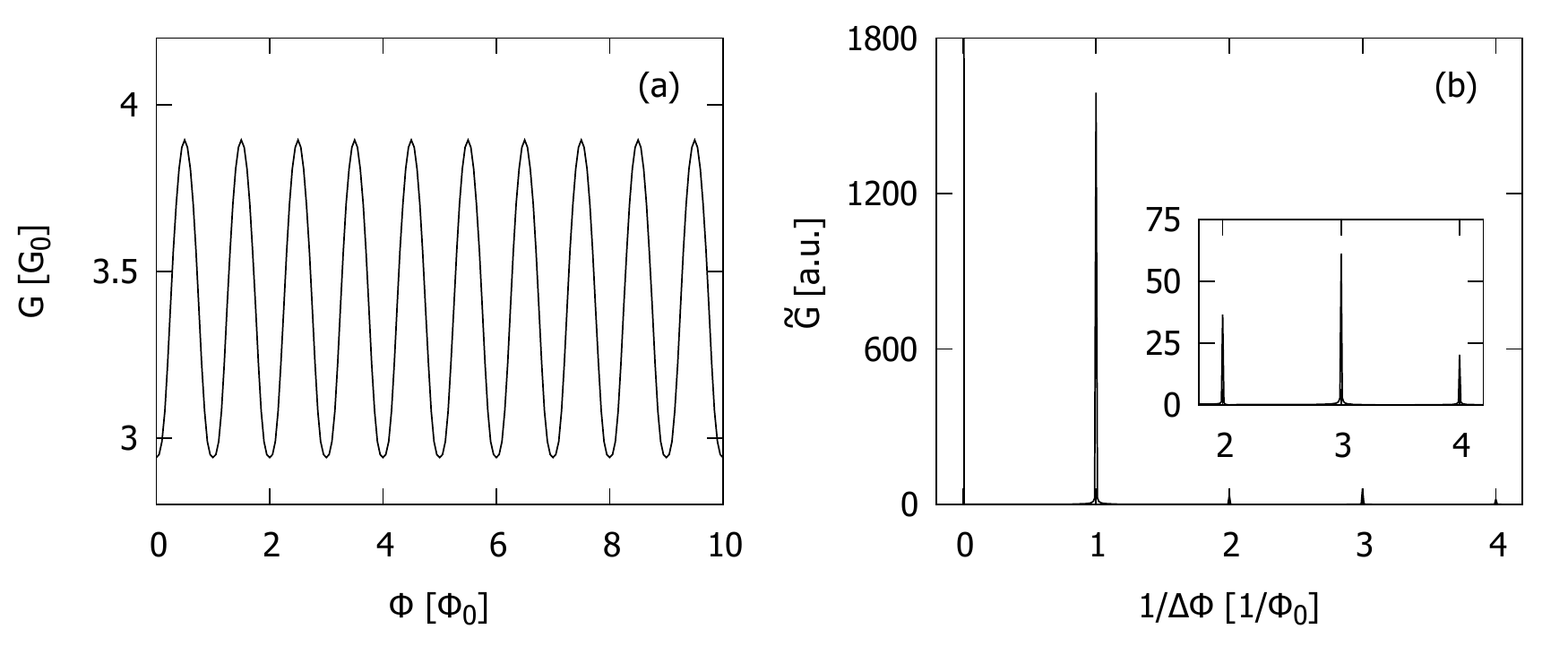}
\caption{ AB oscillations in the conductance $G (\Phi)$  (a) and corresponding power spectrum $\tilde{G} (1/\Delta\Phi)$ (b)  
for the setup of Fig.~\ref{fig1} with $E/t_{0}=0.5$, $B_{z}=\Phi \,\delta\bk{x}\delta\bk{y}$ and  $V_0=\Delta^{(b,e)}=0$. }
\label{fig2}
\end{figure}

We first consider a magnetic field ${\bf B}=(0,0,\Phi\,\delta\bk{x} \delta \bk{y})$, piercing through the ring while being finite on the $z$-axis only. Then the electrons circulating through the system will feel solely the vector potential ${\bf A}$, not a Lorentz force $\propto{\bf B}$. Using the vector potential $\bvec{A}=\Phi \,\theta \bk{-y} \delta \bk{x} \bvec{e}_{x}$ for ${\bf B}$,  the enclosed flux $\Phi=\oint \boldsymbol{A} \text{d}\boldsymbol{r}$ is the same 
for all closed electron paths on the ring, which means,  in the numerical analysis, the total Peierls phase shift $ 2\pi  \Phi/\Phi_{0}$ can simply be assigned to the  electron after passing the angle $\varphi=3\pi /2$ (instead of calculating all  Peierls phases between consecutive pairs of sites $i$ and $j$ along the actual path). Accordingly, simple cosine AB oscillations develop, see Fig.~\ref{fig2}~(a). The corresponding Fourier spectrum $\tilde{G} (1/\Delta\Phi)$ reveals, besides the constant offset at $1/\Delta \Phi=0$, a pronounced peak at the fundamental AB frequency $1/\Delta \Phi=1/\Phi_0$. Caused by multiple cycles of the electrons, the peaks of the higher harmonics have much smaller spectral weight [cf. the inset of Fig.~\ref{fig2}~(b)].

\begin{figure}[b!]
\includegraphics[scale=0.66]{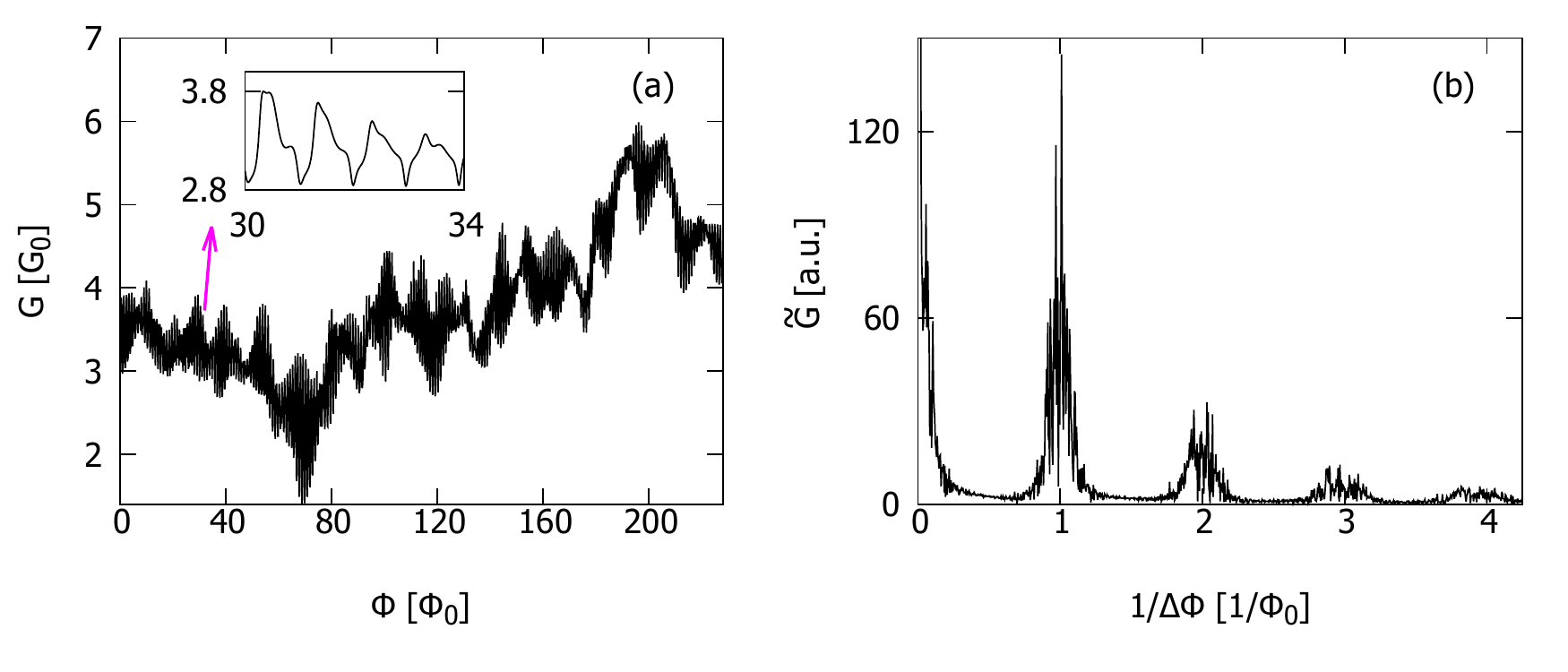}
\caption{(a) Magneto-conductance $G$ as a function of the flux $\Phi= B \pi  \bar{r}^2$  applying a homogeneous magnetic field $B_{z}=B$. Again,  $E/t_{0}=0.5$, and $V_0=\Delta^{(b,e)}=0$. The inset  clearly resolves the AB oscillations. (b) Corresponding Fourier spectrum $\tilde{G} (1/\Delta\Phi)$. The broadening of the peak at $1/\Delta \Phi=0$ causes the low-frequency conductance fluctuations in (a), being absent in Fig.~\ref{fig2}.}
\label{fig3}
\end{figure} 

Next, we consider a  homogeneous magnetic field  ${\bf B}=(0,0,B)$ that is finite in the entire space, and therefore influences the electrons also directly by the Lorentz force. The inset in Fig.~\ref{fig3} demonstrates that AB oscillations will still occur, on top of (low-frequency) universal conductance fluctuations, but are reduced in amplitude and have no longer a simple cosine form. Here we defined $\Phi= B \pi  \bar{r}^2$ as the flux through a circle with average ring radius $\bar{r}=(r_1+r_2)/2$~\cite{WW10}. Although  the peaks in the power spectrum lie very close to multiples of $1/\Phi_0$, they are slightly shifted and broadened in view of the finite width of the ring. Now the electrons affected by the Lorentz force will circulate on different routes, thereby picking up fluxes that vary in strength. The increased weight of the higher harmonics indicates that the probability of multiple cycles is markedly enhanced (compare the corresponding intensities in Figs.~\ref{fig2} and~\ref{fig3}).

\begin{figure}[b]%
\includegraphics[width=0.97\textwidth]{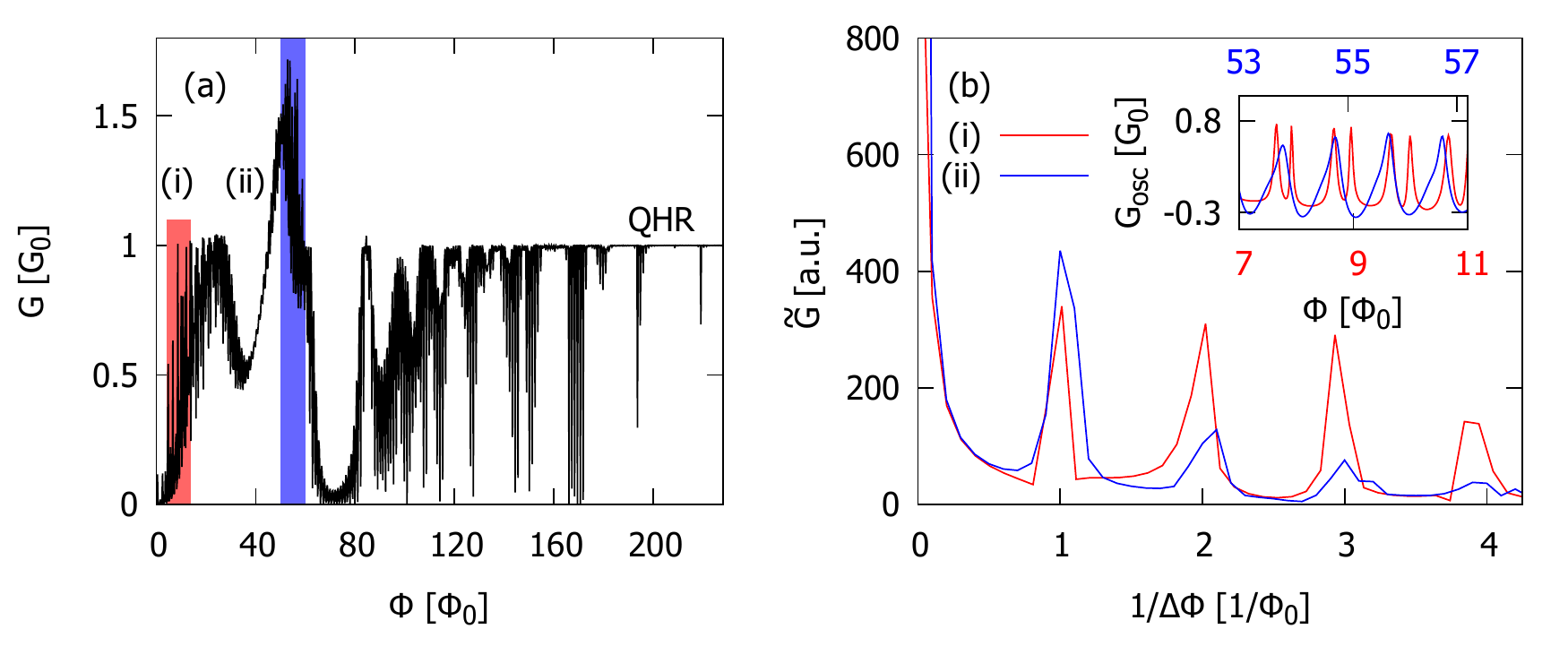}
\hspace*{0.9cm}
\includegraphics[width=0.9\textwidth]{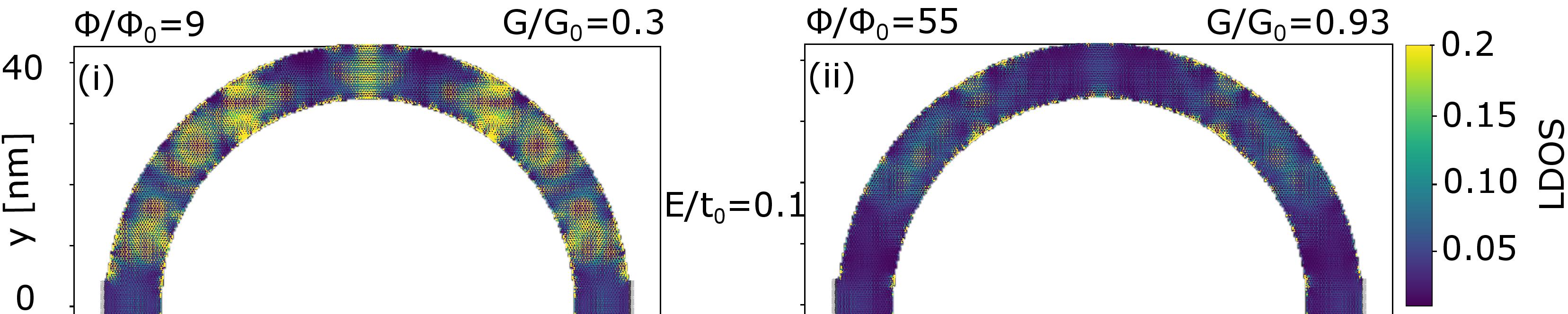}
\caption{Upper part: (a) Dependence of magneto-conductance $G$ on $\Phi= B \pi  \bar{r}^2$ ($B=B_z=$ const.) for $E/t_{0}=0.1$, $V_0=0$ and $\Delta^{(b,e)}=0$. Marked in red and blue are the cases (i) $\Phi\approx 9 \Phi_{0}$ and (ii) $\Phi\approx 55 \Phi_{0}$, respectively, for which the Fourier spectra $\tilde{G} (1/\Delta\Phi)$, calculated within the intervals $\left[4,\ldots,14\right]\Phi_{0}$ and $\left[50,\ldots, 60\right]\Phi_{0}$, respectively, are depicted in (b). In (b) the inset gives the oscillating part of the conductance, $G_{osc}$, which is obtained subtracting the universal conductance fluctuations by a high-pass filter with cut-off frequency $1/(2\Phi_0)$. Lower part: LDOS for cases (i) and (ii) (here only the upper arm of the ring is shown for symmetry reasons). Note that the mean LDOS (conductance) is by a factor of three larger (smaller) for $\Phi\approx 9 \Phi_{0}$ [left panel] if compared to those for  $\Phi\approx 55 \Phi_{0}$  [right panel], indicating particle ``trapping'' in the ring (lower transmittance of the ring structure).}
\label{fig4}
\end{figure}

\subsection{Aharonov-Bohm effect versus Quantum Hall effect}
We now investigate the behavior of the magneto-conductance for a wider range of fluxes, up to magnetic fields that realize the quantum Hall regime (QHR) in the sample. To understand the obtained results, we recall that the Lorentz force (acting now in the ring arms for the homogeneous magnetic field configuration considered), forces the electrons --   within a semiclassical picture -- on a cyclotron path with diameter $d_{c}=2E/veB$, where $v =3t_{0}a_0/2 \hbar$ is the Fermi velocity (in bulk graphene).  Figure~\ref{fig4}, showing the conductance as a function of the average flux for $E/t_0=0.1$, indicates  different regimes.
At low and moderate fluxes the electron current almost equally flows through both arms. That is why we observe pronounced AB oscillations.  Thereby, for $\Phi \simeq 9 \,\Phi_{0}$ [case (i)], the cyclotron diameter $d_c$ matches the average ring diameter  $(r_1+r_2)$ and, as a result, the probability is enhanced for electrons to circulate several times through the system before  leaving the ring interferometer. This is confirmed by the Fourier spectrum presented in Fig.~\ref{fig4} (b) (red curves), as well as by the LDOS  which indicates an increased particle confinement in the graphene ring triggered by the cyclotron dynamics (see left lower panel). In case that $d_c$ will not match the ring diameter, such as for $\Phi \simeq 55 \,\Phi_{0}$  [case (ii)], the higher harmonics in the power spectrum are much less pronounced and the particle will more readily leave the ring after one pass (cf. also the corresponding LDOS displayed in the right lower panel). That the higher harmonics play a prominent [an almost negligible] role in case (i) [case (ii)] becomes apparent when   the universal conductance fluctuations are subtracted from $G$: After that the overtones become clearly visible in the cyclotron regime (see inset of  Fig.~\ref{fig4} (b), where the oscillating part of the conductance, $G_{osc}$, is given). 

For even larger magnetic fields, the cyclotron diameter becomes comparable to the ring arm width $(r_2-r_1)$. 
In this case, the electrons are more and more forced to propagate solely through the upper arm of the graphene interferometer  until, at very high fluxes, interference can no longer take place, i.e., the AB oscillations disappear.  Instead the conductance becomes almost constant,  $G\simeq G_{0}$, which indicates that the system enters the QHR where quantum Hall edge channels evolve. This happens, for our ring sizes, if $\Phi > 80 \Phi_0$ (at $E/t_{0}=0.1$), see the plateau in Fig.~\ref{fig4}~(a). It has been pointed out that very small AB oscillations might nevertheless occur due to interference between paths going along the upper arm and leaving the ring after a half circle and paths circulating one more time around the ring~\cite{WW10}. This will happen only if the width of the lead is somewhat smaller than the width of the ring, as then quantum Hall edge channels of one arm will be scattered at the opening of the leads and therefore may enter the other arm~\cite{WW10}. 

We thus finally conclude that the (topological) AB effect was here observed for a homogeneous magnetic field that penetrates the whole
graphene ring-shaped structure, not only the hole in the middle. This holds at least for rings with arm widths small compared to the ring diameter (in our case, we had $(r_2-r_1)/(r_2+r_1) \simeq 0.1$), and magnetic fields below the QHR.        

\subsection{Disorder effects}
\label{disorder}
Disorder naturally affects the transport properties of contacted graphene nanostructures~\cite{PSWF13}. Even Anderson localization~\cite{An58} occurs in disordered graphene nanoribbons. However, taking into account the large localization length for weak disorder in two or quasi-one dimensions, compared with the device dimensions,  most actual systems appear to be conducting~\cite{SSF09,SSF10}.

To investigate the impact of disorder on the AB effect in graphene rings we consider the setup of Fig.~\ref{fig1} and assume  that particle transmission is influenced by random on-site potentials on the lower arm of the interferometer.  For a direct comparison with the results obtained for a clean sample, we calculate the conductance for bulk disorder $\Delta^{(b)}$ of strength $\gamma_b/t_0=1$, whereby the other parameters  are the same as in Fig.~\ref{fig3}. We have to remark that all   results presented for disordered systems in what follows belong to a single (but typical) sample. 

\begin{figure}[t]%
\centering
\includegraphics[width=1\textwidth]{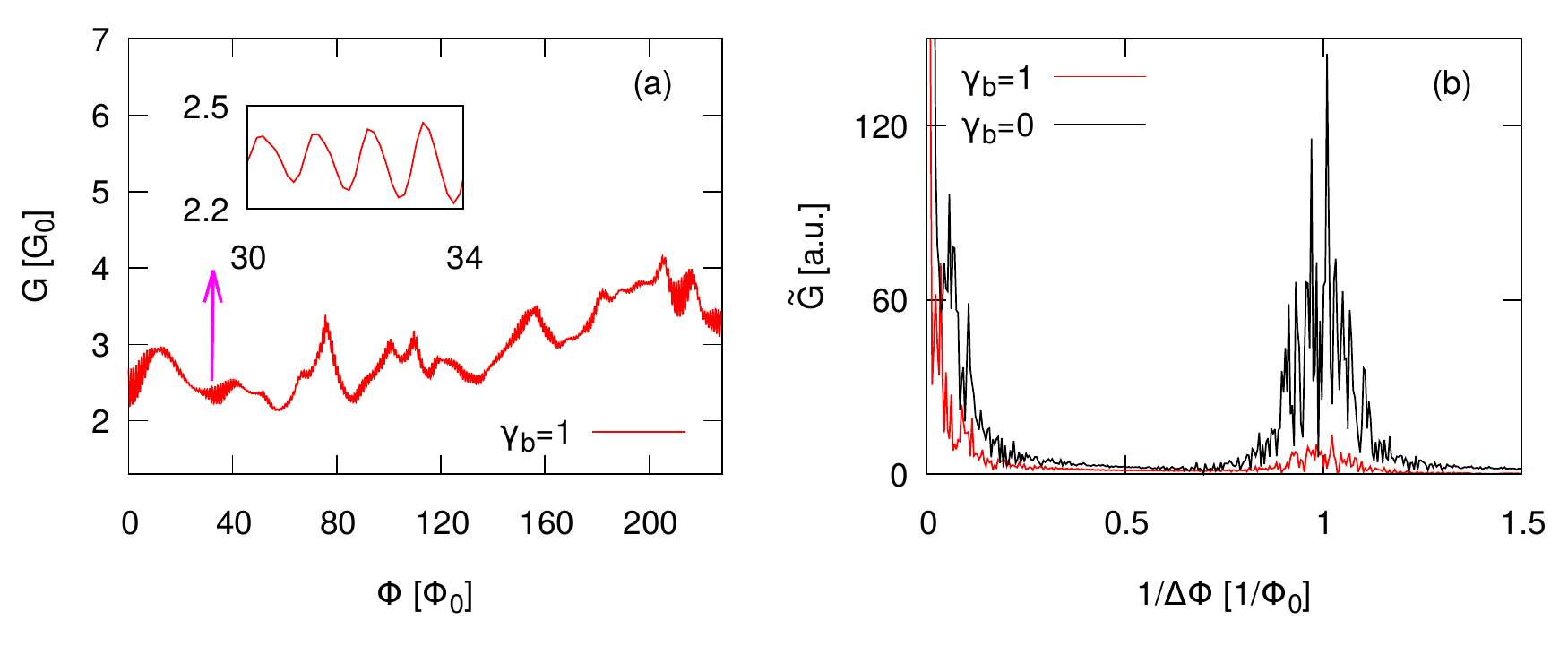}
\caption{(a) Magneto-conductance $G(\Phi)$ and (b) Fourier spectrum $\tilde{G} (1/\Delta\Phi)$  for a graphene interferometer with bulk disorder in the  lower arm ($\gamma_b/t_0=1$). We use $E/t_0=0.5$, $V_0=0$, $\Delta^{(e)}=0$, and apply  a constant perpendicular magnetic field $B=B_z=$ const. The inset in (a) enlarges the weak AB oscillations. The black curve in (b) gives $\tilde{G} (1/\Delta\Phi)$ without disorder  ($\gamma_b/t_0=0$).}
\label{fig5}
\end{figure}

Figure~\ref{fig5} demonstrates the drastic reduction of the amplitude of the AB oscillations as a consequence of the impurity scattering  in the lower arm. Clearly the conductance as a whole and the universal conductance fluctuations will be also reduced (compare the magnitudes  of $G/G_0$ in Figs.~\ref{fig3} and~\ref{fig5}). Most pronounced, however,   is the reduction of the spectral weight of the higher harmonics in the Fourier spectrum.
\begin{figure}[htb]%
\centering
\includegraphics[width=1\textwidth]{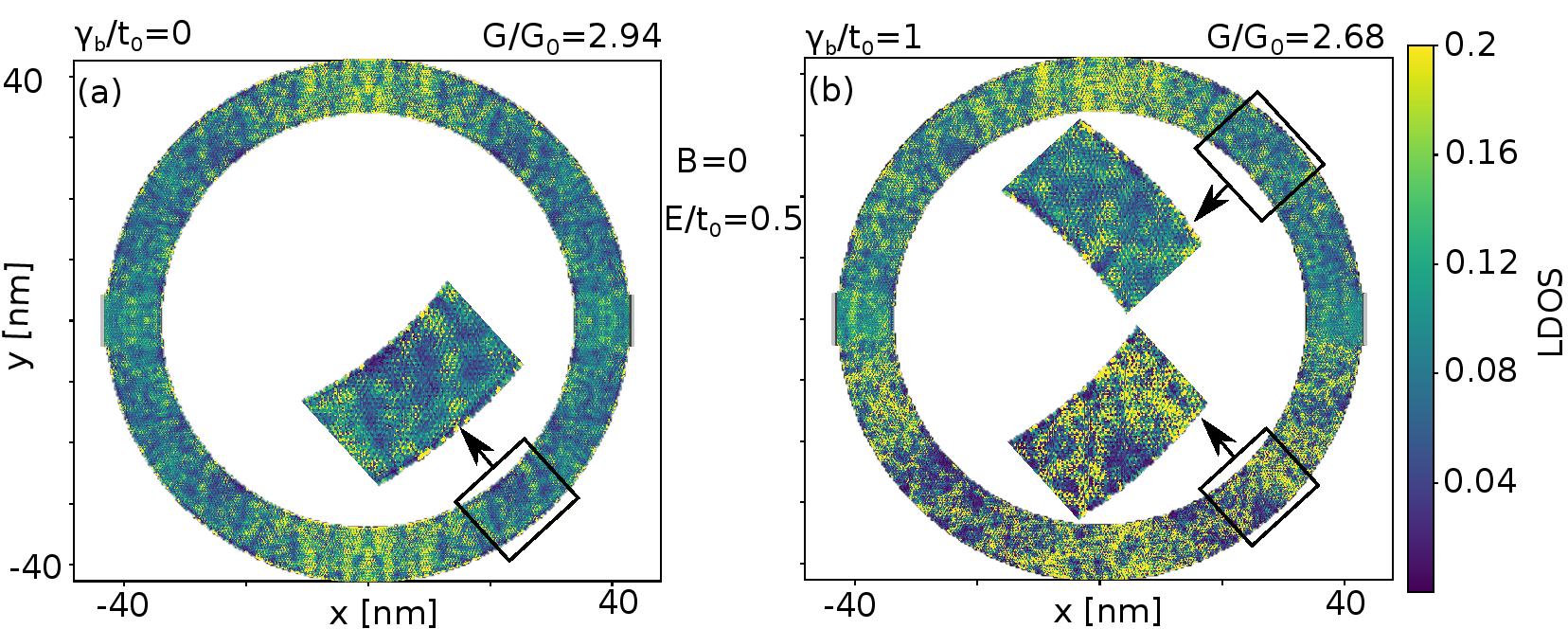}
\caption{Local density of states  for a graphene ring without (left)  and with bulk disorder (right) in the lower arm. Parameters are the same as in Fig.~\ref{fig5}.} 
\label{fig6}
\end{figure}

Further information about the electronic structure of the system can be obtained from the LDOS depicted in Fig.~\ref{fig6}.
Of course, the LDOS of the upper and lower arms are related in mirrored symmetry  for $\gamma_b=0$ (left panel).   Because of the contacts and the (irregular) edges of the ring we observe a spatial variation of the LDOS even in this case.
The LDOS mirror symmetry between the upper and lower arms is broken for the disordered sample. Now the electron wave packet  is strongly scattered (temporarily trapped and re-emitted) by the impurity potentials in the lower arm. This results in a much stronger spatial variation of the LDOS. The magnification reveals filamentary and puddle-like structures in the lower arm with density fluctuations of at least one order of magnitude on length scales of a few nanometers, i.e., much larger than  the carbon-carbon distance (see right panel). These puddle-like structures, resulting from subtle interference quantum effects, lead to dominant intra- and inter-puddle transport and thereby mask Anderson localization~\cite{SF12a}. A more thorough investigation of the localization properties would require the calculation and analysis of the distribution of the LDOS~\cite{SSBFV10}, which is beyond the scope of this short contribution however.

\subsection{Observation of Klein tunneling with a gated Aharonov-Bohm interferometer}
\label{edge_disorder}
A precondition for the observation of Aharonov Bohm oscillations is the quantum interference between electrons that have passed the ring on different paths, i.e., a non-vanishing transmittance in both arms of the ring. Remarkably a sharp and high ($V_0>E$ ) potential barrier, created on the lower arm by a local gate, should not affect the AB oscillations by reason of Klein tunneling  that enables an unimpeded transmission of electrons through such a $npn$-junction type system. While Klein tunneling strongly  depends on the smoothness of the potential, the details of the gate potential interface 
are almost unimportant for normal tunneling  processes (without Klein tunneling only evanescent waves will appear in the tunnel process). This allows to detect Klein tunneling by analyzing the transmission through a smooth potential~\cite{SB10}, as the one in Eq.~(\ref{potential}). Here, in addition, we study the interplay of Klein tunneling and disorder by taking into account random on-site potentials in~Eq.~(\ref{eq1}).

In order to measure the strength of the AB oscillations we exploit, following~\cite{SB10}, the root mean square (RMS) amplitude of the conductance,
\begin{align} 
\Delta G_{RMS}=\sqrt{\frac{1}{M}\sum_{j}^{M}G^2_{j}}\;,
\end{align}
where $M$ is the number of conductance points  $G_j$ calculated in a certain flux range. In our case, we use $M=1001$, $\Phi=[0...228] \Phi_{0}$, and a high pass frequency filter with cut-off frequency $1/(2\Phi_0)$ to subtract the background of universal conductance fluctuations. 

 \begin{figure}
\centering
\includegraphics[width=1\textwidth]{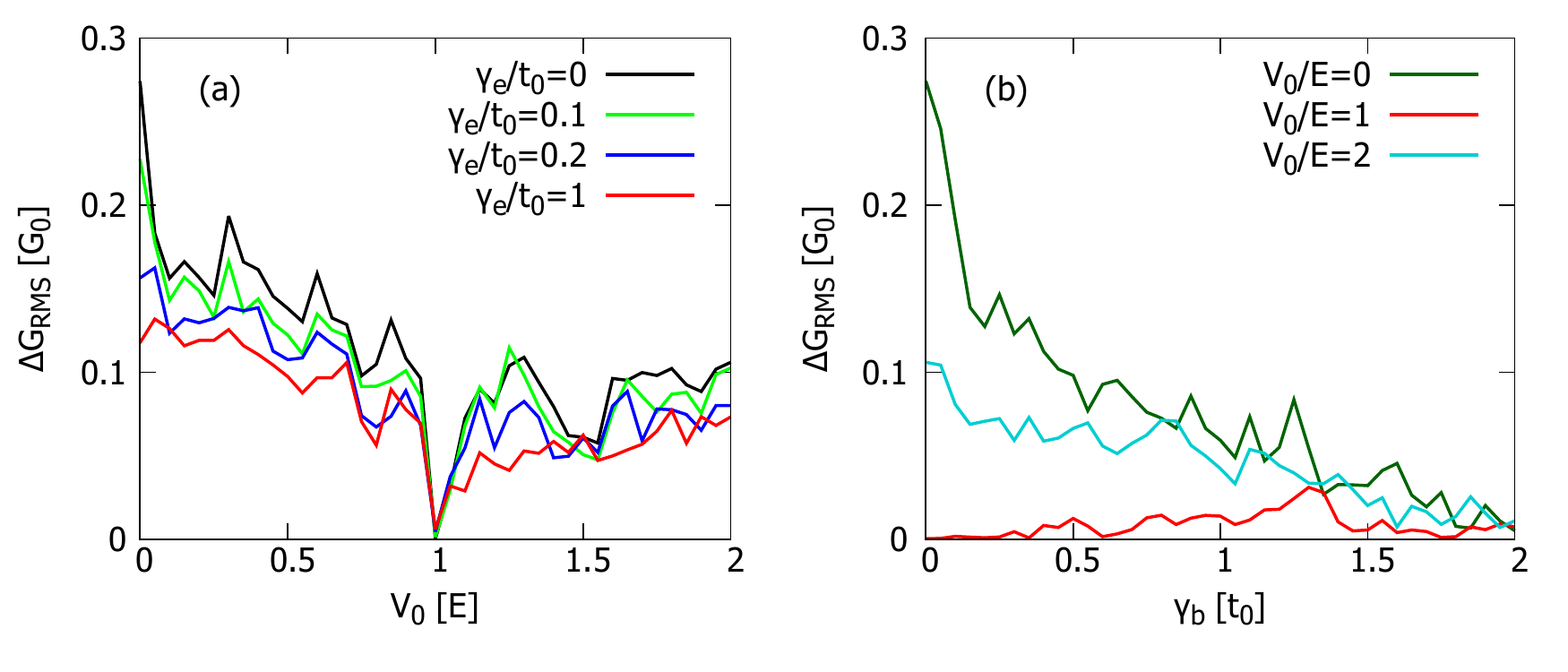}
\caption{(a) RMS magnitude of the conductance $\Delta G_{RMS}/G_0$ as a function of   the relative height of the potential barrier $V_{0}/E$  for different edge disorder strengths $\gamma_{e}$. (b) $\Delta G_{RMS}/G_0$ in dependence on the bulk disorder strength $\gamma_{b}$ for $V_0/E=0$ ($n$ band),  $V_{0}/E=1$ (charge neutrality point), and $V_{0}/E=2$ ($p$ band). Again, we use $E/t_0=0.5$ and $B=B_z=$ const.}
\label{fig7}
\end{figure}

Figure~\ref{fig7} (a) shows the dependence of $\Delta G_{\text{RMS}}$ on the relative potential height $V_0/E$. 
For $V_{0}/E<1$ ($V_{0}/E>1$) the lower arm of the graphene ring forms a $nn'n$- ($npn$-) junction, while there 
are no transport channels at the charge neutrality point  $V_{0}/E=1$. Neglecting disorder effects 
($\gamma_{b,e}=0$), we see that the RMS magnitude of the AB signal is reduced if the potential barrier gets higher. This is because the potential scattering increases  and, at the same time, fewer and fewer transport channels act to transmit the particle through the $nn'n$-junction. At $V_{0}/E=1$, in the lower arm, we have no states (transport channels) at all, which means that AB interference cannot be realized.  Increasing $V/E$ further, an $npn$-junction forms, and Klein tunneling becomes possible. Then transmission through the lower arm happens again and can be viewed, in a certain sense, as (coherent) hole propagation, until the particle pops up as an electron when leaving the barrier.  As a result, the AB oscillations reappear and increase in magnitude raising $V_0/E$, simply because  the number of transport channels increases. Because our potential is rather smooth, the Klein tunneling through the lower arm will not be perfect (there are spatial zones where the energy $E$ of the  wave packet is comparable to $V$). Consequently, the AB oscillations will be reduced in magnitude if compared to the case $V_0/E<1$. Most notably, however, Klein tunneling can be observed in principle with such a setup.  Additional edge disorder somewhat suppresses the AB oscillations (we emphasized already that the edges of the ring are irregular in any case), but the general curve characteristics remain unchanged.
 
 Bulk disorder notably weakens  the AB effect. This can be clearly seen from Fig.~\ref{fig7} (b), showing the value of $\Delta G_{\text{RMS}}$ for increasing $\gamma_b$ in the $nn'n$- (green curve) and $npn$-  (blue curve) junction regimes. Interestingly, for $V_0/E=1$, i.e., at the charge neutrality point, disorder of  adequate (average) strength might even promote particle transmission through the lower arm by impurity scattering, and therefore also gives rise to AB oscillations. Of course, very strong bulk disorder will lead to (Anderson) localization and thereby  suppresses the AB effect.
 
\section{Conclusions}
\label{conclus}
To summarize, we have investigated a contacted graphene ring exposed to magnetic and electrostatic fields, which, as a matter of principle, embodies an graphene-based Aharonov-Bohm interferometer with local gates. Simulating the transport properties of such a mesoscopic device, bulk and edge disorder, realized in the sample by random on-site potentials, was     taken into account as well. By means of exact numerical techniques we solved the scattering problem of an incoming Dirac electron wave packet within the Landauer-B\"uttiker scheme and, in the main, analyzed the magneto-conductance. Focusing on the interplay of the Aharonv-Bohm effect, normal or Klein tunneling, and localization effects we examined, corroborated and extended previous theoretical findings for similar setups~\cite{Reea07,SZ16,JMPT09,SB10,WW10,KKM17,SRT12}.
In particular, we verified that the AB oscillations, appearing on top of universal conductance fluctuations, will be suppressed when very large fields (fluxes) penetrate the sample and the system enters the quantum Hall regime.  Disorder greatly reduces the conductance of the graphene ring and also and especially the magnitude of  the AB oscillations. For strong disorder the local density of states even indicates particle localization effects, but Anderson localization in a strict sense seems to be prevented by the formation of spatial charge imbalances, e.g., electron-hole puddles, in such mesoscopic devices. Near the charge neutrality point, however, disorder might induce transport and thereby revive the AB effect.  We furthermore show that with local electrostatic  gates imprinting smooth potential barriers  the Klein tunneling phenomenon can be influenced in a controlled way.  

While the AB effect in graphene rings has already been observed experimentally~\cite{RO08,NKPKKL11,SSH12}, the proposed setup seems to be promising to detect also Klein tunneling in future experiments~\cite{SRT12}, whereby the disorder effects we have discussed will be of special importance.
\begin{acknowledgements} 
This work was supported by Deutsche Forschungsgemeinschaft through the Collaborative Research Center SFB 652 (project B5) and the Competence Network for Scientific High-Performance Computing in Bavaria (KONWIHR III, project PVSC-TM). HF acknowledges the hospitality at the Los Alamos National Laboratory where part of this work was performed.
\end{acknowledgements}


\end{document}